\documentclass[final,5p,times,twocolumn]{elsarticle}
\biboptions{sort&compress} 

\usepackage{amssymb}
\usepackage{amsmath}
\usepackage{cuted} 
\usepackage{comment}
\usepackage{url}

\usepackage{hyperref}
\usepackage[capitalise]{cleveref}
\usepackage{xcolor}
\usepackage{upgreek} 
\biboptions{square} 
\definecolor{nicegreen}{rgb}{.1,.5,.1}
\definecolor{darkblue}{rgb}{0,0,.5}
\hypersetup{colorlinks,
    }

\usepackage{commands} 

\usepackage{scalerel}
\usepackage{atbegshi}
\makeatletter
\AtBeginShipout{%
  \ifnum\value{page}=1\relax
    \setbox\AtBeginShipoutBox=\hbox to \wd\AtBeginShipoutBox{%
      \kern 3mm%
      \hbox to \dimexpr\wd\AtBeginShipoutBox-3mm\relax{%
        \box\AtBeginShipoutBox\hss
      }%
    }%
  \fi
}
\makeatother
\journal{Physics Letters B}

\begin{document}

\begin{frontmatter}

\title{Leaking Sound Waves: Gravitational Waves From Dark Sectors Weakly Coupled to the Standard Model}

\author[UA,UM,first]{Marco Finetti}
\author[first]{Helena Kole\v{s}ov\'{a}}
\author[first]{Daniil Krichevskiy}

\affiliation[first]{organization={University of Stavanger, Department of Mathematics and Physics}, 
            addressline={Kristine Bonnevies vei 22}, 
            city={Stavanger},
            postcode={4021}, 
            country={Norway}}
\affiliation[UA]{
    organization={Departamento de Física, Universidade de Aveiro},
    addressline={Campus de Santiago}, 
    city={Aveiro},
    postcode={3810-183}, 
    country={Portugal},
    }
\affiliation[UM]{
    organization={Laboratório de Instrumentação e Física Experimental de Partículas (LIP), Universidade do Minho},
    city={Braga},
    postcode={4710-057}, 
    country={Portugal},
    }

\begin{abstract}
First-order phase transitions in a dark sector can produce an observable stochastic gravitational-wave background, and many models predicting such signals in current and future experiments have been explored. 
For moderately strong transitions, hydrodynamical simulations suggest that sound waves generated by the expansion and collisions of true-vacuum bubbles are the dominant source of gravitational waves. It is often assumed that these sound waves propagate only within the dark-sector fluid, while the impact of the visible sector is neglected. In this paper, we investigate the effects of a weak coupling between the dark and visible fluids on the resulting gravitational-wave signal. We adopt a phenomenological hydrodynamical description of two ultra-relativistic fluids, parametrizing their interactions without specifying an underlying microscopic model. We focus on the regime in which the coupling can be neglected during the phase transition but becomes relevant during the subsequent, longer-lasting period of sound wave propagation. Within this framework, we analytically show that, even if the bubble expansion initially affects only the dark fluid, sound waves can subsequently be induced in the visible sector.
Using the sound-shell model, we find that, despite the additional sound waves generated in the visible sector, the total gravitational-wave signal sourced by both fluids is generally suppressed compared to the case of a completely decoupled dark sector. This suppression becomes more pronounced for larger interaction rates and a larger ratio of relativistic degrees of freedom in the visible sector to those in the dark sector. The coupling also modifies the shape of the gravitational-wave spectrum.
\end{abstract}


\end{frontmatter}

\section{Introduction} 
\label{introduction}

Recent advances in gravitational-wave (GW) astronomy give hope that the early Universe can also be probed through this new observational window. Indeed, violent processes like a cosmological first-order phase transition may have left traces in the form of a GW background, and such a signal might be observable by future experiments like LISA~\cite{Caprini:2024hue}. Moreover, a stochastic GW background was recently observed by pulsar-timing arrays (PTAs)~\cite{NANOGrav:2020bcs,NANOGrav:2023gor} and although this signal may come from astrophysical sources, one of the alternative explanations is a cosmological first-order phase transition~\cite{Nakai:2020oit,Ratzinger:2020koh,NANOGrav:2021flc,Bringmann:2023opz,Han:2023olf,Madge:2023dxc,DiBari:2023upq,NANOGrav:2023hvm,Winkler:2024olr,Banik:2024zwj,Costa:2025csj,Goncalves:2025uwh,Bringmann:2026xcx,Musumeci:2026jum}. In the Standard Model (SM) of particle physics, no first-order phase transition is expected~\cite{Kajantie:1996mn,Aoki:2006we}, 
hence, a dark sector with new scalar fields acquiring a vacuum expectation value is usually devised to explain the PTA data~\cite{Nakai:2020oit, Bringmann:2023opz, Han:2023olf,DiBari:2023upq, Banik:2024zwj,Costa:2025csj,Goncalves:2025uwh,Musumeci:2026jum}.
In general, cosmological phase transitions in dark sectors were considered in many works (see~\cite{Breitbach:2018ddu, Ertas:2021xeh, Bringmann:2023iuz, Biondini:2026uds} for further examples); interestingly, dark sectors may feature phenomena not present in the SM like a first-order confinement or chiral phase transition~\cite{Schwaller:2015tja,Huang:2020crf,Reichert:2021cvs,Morgante:2022zvc,Pasechnik:2023hwv}.

The works mentioned above considered one of the two limiting cases of either the dark sector being fully secluded or forming a single fluid with the SM plasma. The reason for selecting one of these regimes was often not discussed in detail, although this choice is important for the evaluation of the GW spectrum. Indeed, the anisotropic stresses that induce GW production arise from the fluid motion triggered by the phase transition, and it is important to determine whether only the dark sector fluid is perturbed or if the perturbations extend to the SM fluid as well. In this work, we discuss qualitative distinctions among the various coupling regimes between the dark and visible sectors, and study in detail an intermediate regime where this coupling affects only the late stage of the GW production, characterized by the sound wave propagation. Using the analytic sound-shell model~\cite{Hindmarsh:2016lnk,Hindmarsh:2019phv,RoperPol:2023dzg}, we show that the coupling between the dark and visible sectors leads to a suppression of the GW signal compared to the case of a secluded dark sector. 

This work is structured as follows. 
In \cref{sec:DarkSectors}, we discuss the criteria used to distinguish the different regimes of dark sector coupling to the visible fluid relevant to first-order phase transitions and introduce a new regime of weakly coupled dark sectors.
In \cref{secHydrodynamics}, we present the hydrodynamic modeling of the energy and momentum exchange between the dark and visible fluids and derive the linearized equations governing the fluid perturbations. Their solutions are used in \cref{sec:GW} to calculate the GW spectrum, which is then evaluated numerically in \cref{sec:numerics}. Finally, we summarize our results and outline directions for future research in \cref{sec:conclusions}.

\section{Secluded or coupled dark sector?}\label{sec:DarkSectors}
To understand the possible effects of a coupling between the dark and visible fluids, let us first review the mechanism of GW production from cosmological first-order phase transitions. 
Such phase transitions typically proceed via the nucleation of bubbles containing the new vacuum. Consequently, part of the released latent heat $\epsilon$ is transformed into bulk fluid motion.
Hydrodynamic simulations show that, for moderate phase transitions, this fluid motion represents the main source of GWs since it lasts much longer than the phase transition itself. In particular, we concentrate here on the sound waves propagating in the plasma after the phase transition completion, and the duration of this source is usually estimated as the time before nonlinearities in the fluid develop: 
$\tau_{\rm sw} \sim \tau_{\rm nl}\sim R_\ast/\sqrt{K}$ where $R_\ast$ is the mean bubble separation and $K$ denotes the fraction of the total fluid energy stored in bulk fluid motion with typical values $K \sim \mathcal{O}(10^{-2})$~\cite{Caprini:2019egz,Caprini:2024hue}.
On the other hand, the phase transition duration $1/\beta$ is given by $R_\ast$ up to an $\mathcal{O}(1)$ factor: $1/\beta = R_\ast / (\sqrt[3]{8\pi}\, v_w)$ where $v_w$ is the bubble wall velocity. Consequently, the period of sound wave propagation $\tau_{\sw}$ extends over timescales significantly longer than the phase transition duration.

It is expected that if the timescale of interactions between the dark and visible fluids $\tau_\inter$ is much longer than $\tau_\sw$, then only the dark fluid will be perturbed by the phase transition dynamics, hence, the visible fluid will not contribute to GW production. Consequently, for this case of a secluded dark sector, one of the key quantities determining the GW amplitude, namely, the factor $\kappa$ quantifying the efficiency of the vacuum energy transfer to the bulk fluid motion~\cite{Espinosa:2010hh} is evaluated based on the dark sector properties only. In particular, $\kappa$ depends on the phase transition parameter $\alpha_d$, which we evaluate assuming the bag equation of state as in~\cite{Espinosa:2010hh}:
\begin{equation}\label{eq:alphad}
    \alpha_d \equiv \frac{\epsilon}{\bar{\rho}_{d,\sym} - \epsilon}\,,\quad \bar{\rho}_{d,\sym} - \epsilon = \frac{\pi^2}{30}\, g_{d,\sym} T^4\,.
\end{equation}
Here $g_{d,\sym}$ is the effective number of the relativistic degrees of freedom in the dark sector's symmetric phase and $T$ is the temperature.\footnote{In this work, we consider single temperature $T$ shared by the dark and visible sectors. For very weak interactions ($\tau_\inter \gtrsim 1/H_\ast$), the dark and visible sector temperatures may differ, which has further implications for the GW spectrum, see~\cite{Breitbach:2018ddu, Ertas:2021xeh}. }

If, on the other hand, the interactions are fast compared to the phase transition duration $\tau_\inter \ll 1/\beta$, both dark and visible fluids are perturbed already at the stage of bubble nucleation. In this regime, the dark sector forms a single fluid with the visible plasma as assumed, e.g., in~\cite{Balan:2025uke}. The efficiency factor $\kappa$ is then calculated using 
\begin{equation}\label{eq:alphatot}
    \alpha_\tot \equiv \frac{\epsilon}{\bar{\rho}_{\tot,\sym}-\epsilon}\,,\quad \bar{\rho}_{\tot,\sym} - \epsilon = \frac{\pi^2}{30}\, g_{\tot,\sym} T^4
\end{equation}
where $g_{\tot,\sym} = g_{d,\sym} + g_v$ includes the effective number of relativistic degrees of freedom in the visible sector $g_v$. This contribution might be large: $g_{v} \approx 100$ at temperatures $T\gtrsim 100\,$GeV and $g_{v} \approx 10$ at temperatures $T\sim  10\,$MeV relevant for the PTA signal. On the other hand, minimal dark sectors that allow for a first-order phase transition may contain very few effective relativistic degrees of freedom, e.g., $g_{d,\sym} = 4$ for the Abelian dark Higgs model~\cite{Biondini:2026uds}. Consequently, $\alpha_\tot$ might be substantially suppressed compared to $\alpha_d$, leading to considerable suppression of the GW signal as observed, e.g., in~\cite{Biondini:2026uds}.

Motivated by the impact of the coupling between the dark and visible sectors on the GW spectrum explained above, we study in this work an intermediate regime in which the bubble nucleation and growth affect the dark fluid only, but the visible fluid is still perturbed later, during the period of sound wave propagation. This regime is expected to appear if
\begin{equation}\label{eq:Timescales}
    \beta^{-1} \ll \tau_\inter \lesssim \tau_\sw.
\end{equation}
Although the value of $\tau_\inter$ (or equivalently the corresponding interaction rate $\Gamma_\inter \sim 1/\tau_\inter$) can, in principle, be calculated from the microscopic structure of a given model, we treat it as a free parameter in this work. Let us note that values of the interaction rates satisfying \cref{eq:Timescales} were considered, e.g., in~\cite{Biondini:2026uds}. Further motivation can be found in Ref.~\cite{Bringmann:2023opz}, 
pointing out that the dark sector particles should decay fast enough 
in order to obtain an explanation of the PTA data compatible with the cosmological constraints~\cite{Yeh:2022heq}. For a subset of the parameter space explored in Ref.~\cite{Bringmann:2023opz}, the dark particles' lifetimes also satisfy \cref{eq:Timescales}.

\section{Two-fluid hydrodynamics
}\label{secHydrodynamics}

In this section we discuss 
the interactions of dark (d) and visible (v) fluids during the period of sound wave propagation using the hydrodynamic model of two weakly-coupled fluids against the expanding FRW background.  Both fluids are assumed to be perfect, ultra-relativistic, and individually in local thermal equilibrium. Throughout this paper, the indices $\alpha,\alpha' \in \{v,d\}$ label the two fluids.

In conformal coordinates, the FRW metric reads
        $ds^{2}=a^{2}\left( \tau \right)  \left[ -d\tau^{2} +\delta_{ij}   dx^{i}dx^{j}\right]$,
where $a$ is the conformal factor and $\tau$ is the conformal time.
The interaction between the fluids can be captured by the system:
\begin{align}\label{eqInteraction}
    \nabla_{\mu } T^{\mu \nu }_{d}=Q^{\nu },\quad \nabla_{\mu } T^{\mu \nu }_{v}=-Q^{\nu }
\end{align}
where  $Q^\nu$ is a current and the perfect fluid energy-momentum tensor reads
\begin{align}
T_\alpha^{\mu\nu}=w_\alpha u^\mu_\alpha u^\nu_\alpha+p_\alpha g^{\mu\nu}.
\end{align}
Here $u^\mu_\alpha=dx^\mu_\alpha/ds_{\alpha}$ is the fluid four-velocity with proper time $s_{\alpha}$  and the enthalpy density reads 
\begin{align}\label{eq:enthalpy_def}
w_\alpha=\rho_\alpha+p_\alpha = \frac{4}{3}\rho_\alpha=\frac{2\pi^{2} }{45} g_{\alpha}  T_\alpha^{4},
\end{align}
where $g_{\alpha}$ is the effective number of relativistic degrees of freedom of the fluid $\alpha$. In particular, the dark sector value in the broken phase $g_{d,\bro }$ will be used in the context of our work.
\cref{eqInteraction} implies that the total energy-momentum tensor is covariantly conserved.
Assuming small temperature and velocity differences between the fluids, both of the same order, Lorentz invariance implies that, to first-order in a near-equilibrium expansion, 
\begin{align}\label{eq:2_fluid_current}
   Q^{\nu }=\varkappa_{\T} (\tilde{T})\left( \frac{T_{v}-T_{d}}{\tilde{T}} \right) \tilde{u}^\nu +\varkappa_{\V} (\tilde{T}) \left( u_{v}-u_{d}\right)^{\nu}.
\end{align}
Here $\varkappa_{\T}$ and $\varkappa_{\V}$ are transport coefficients of mass dimension five, $\tilde T=(T_d+T_v)/2$ and $\tilde u^\mu=(u_d^\mu+u_v^\mu)/2$ denote the mean temperature and four-velocity.\footnote{Since $u_d\!\cdot u_v=-1+\mathcal{O}\!\left((u_d-u_v)^2\right)$, the transport coefficients depend only on $\tilde T$ to linear order.} 
In the absence of any other mass scales, dimensional analysis suggests $\varkappa_{\T/\V}\propto \tilde{T}^5$.
\cref{eqInteraction} with \cref{eq:2_fluid_current} can, in principle, be derived from the first moment of the Boltzmann equation, where the collision integral determines the phenomenological coefficients $\varkappa_{\T/\V}$.

Note that for an ultra-relativistic fluid with the equation of state
$p_\alpha=\rho_\alpha/3$, it holds that  $\nabla_{\mu}T_\alpha^{\mu\nu}=
a^{-6}\partial_{\mu}\!\left(a^{6}T_\alpha^{\mu\nu}\right)$.
Therefore, instead of solving the  hydrodynamic equations in an expanding background, one may equivalently solve the corresponding Minkowski-space equations for the rescaled variables
\begin{align}\label{eqMinkowskiRescaling}
\rho^{\M}_\alpha & = a^4 \rho_\alpha ,\qquad
p^{\M}_\alpha = a^4 p_\alpha, \nonumber \\
u^{\M, \,\mu}_\alpha &= au^\mu_\alpha,\qquad
\varkappa^{\M}_{\T/\V}=a^5\varkappa_{\T/\V},
\end{align}
and recover the physical variables by the inverse rescaling. The time coordinate in the equivalent Minkowski-space equations corresponds to the conformal time $\tau$.  
If $\varkappa_{\T/\V}\propto \tilde{T}^5$, then the coefficients $\varkappa^{\M}_{\T/\V}$ are constant in time because all the $a$-dependence is scaled away. Further, the rescaled velocity reads
\begin{align}\label{eqRescaledVelocity}
 u^{\M, \,\mu}_\alpha &=au^{\mu }_\alpha=a\frac{dx^{\mu }_\alpha}{ds_\alpha }  =\frac{dt}{ds_\alpha } \left( 1,\frac{d\mathbf{x}_\alpha }{d\tau } \right)  \equiv \gamma_\alpha \left( 1,\mathbf{v}_\alpha \right),
\end{align}
where $\mathbf{v_\alpha
}=d\mathbf{x}_\alpha/d\tau$ is  the physical peculiar velocity and $\gamma_\alpha \equiv (1-\mathbf{v}^{2}_{\alpha } )^{-1/2}$. We use the bold face to denote 3-vectors in the whole text.

Linearizing the hydrodynamical equations in small perturbations and velocities, assuming the same background temperature and adiabatic perturbations, i.e., $\delta p^{\M}_\alpha=c_s^2\delta\rho^{\M}_\alpha$, with the same speed of sound $c_s^2=1/3$ in both fluids, we obtain a system of four equations:
\begin{align}\label{wave_equations_eq6}
\begin{cases}\lambda'_{\alpha } +\partial_{i} v^{i}_{\alpha }=\frac{\left( -1\right)^{\delta_{\alpha,d} +1}  }{3\bar{w}_{\alpha } } \varkappa^{\M}_{\T}  \left( \lambda_{v} -\lambda_{d} \right)  ,\\ (v_{\alpha }^{j})' +c^{2}_{s}\partial_{j} \lambda_{\alpha } =\frac{\left( -1\right)^{\delta_{\alpha,d} +1}  }{\bar{w}_{\alpha } } \varkappa^{\M}_{\V}   \left( v_{v}^j-v_{d}^j\right)  ,\end{cases}
\end{align}
where $'\equiv d/d\tau$, $\lambda_\alpha \equiv \delta \rho^{\M}_\alpha/\bar{w}^{\M}_\alpha=\left(\rho^{\M}_{\alpha }-\bar{\rho}^{\M}_{\alpha } \right)  /\bar{w}^{\M}_\alpha\ll1 $, $u_{\alpha}^i \ll 1$ and barred variables denote the background values. Note that the second law of thermodynamics for the total entropy density current,  $\partial_\mu S^\mu\ge0$, requires $\varkappa^{\M}_{\V/\T}\ge0$. 

For a longitudinal flow, the system \cref{wave_equations_eq6} can be conveniently solved by introducing the scalar potentials $\Phi_\alpha$, defined as 
$v_\alpha^i = \partial^i \Phi_\alpha$,
and, correspondingly,
\begin{align}\label{eq:ScalarPotentialAndEnergy}
\lambda_{\alpha}
= -\frac{1}{c_s^2}
\left(
\Phi'_{\alpha}
+\frac{(-1)^{\alpha}}{\bar{w}^{\M}_{\alpha}}
\varkappa^{\M}_{\V}
\left(\Phi_v-\Phi_d\right)
\right).
\end{align}

In Fourier space,\footnote{The Fourier and inverse Fourier transforms are defined as $f\left( \mathbf{k} \right)  =\int \dd^{3}\mathbf{x} e^{i\mathbf{k} \mathbf{x} }f\left( \mathbf{x} \right)$, 
$f\left( \mathbf{x} \right)  =\int \frac{\dd^{3}\mathbf{k} }{(2\pi )^{3}} e^{-i\mathbf{k} \mathbf{x} }f\left( \mathbf{k} \right)$.} the equations for the weighted collective mode $\Phi_{+}=(\bar{w}_{d}\Phi_{d}+\bar{w}_{v}\Phi_{v})/\bar{w}$ (with $\bar{w}
=\bar{w}_d\bar{w}_v/({\bar{w}_d+\bar{w}_v})$) and the relative mode $\Phi_{-}=\Phi_v-\Phi_d$ read
\begin{align}
     &\Phi''_{+}(\tau,\mathbf{k}) +\omega^{2} \Phi_{+}(\tau,\mathbf{k}) =0,\label{eqCollectiveMode}\\
      &\Phi^{\prime \prime }_{-} (\tau ,\mathbf{k} )+\left( \omega^{2} +\frac{\varkappa^{\M}_{\T} \varkappa^{\M}_{\V} }{3(\bar{w}^{\M })^2 } \right)  \Phi_{-} (\tau ,\mathbf{k} )=-2\Gamma_{\text{int} } \Phi^{\prime }_{-} (\tau ,\mathbf{k} )\label{eqRelativeMode},
\end{align}
where $\mathbf{k}$ is the comoving  wave vector, $k\equiv |\mathbf{k}|$, $\omega=c_s k$, $\Gamma_{\text{int} } =\frac{1}{2\bar{w}^{\M} } \left( \frac{\varkappa^{\M}_{\T} }{3} +\varkappa^{\M}_{\V} \right)  $
 and $\bar{w}^{\M} =a^{4}\bar{w} $. Let us note that the subscript $M$ is omitted in ratios invariant under the rescaling \cref{eqMinkowskiRescaling} throughout the text. The system above admits two undamped oscillatory modes,
$\Omega_{+,\pm}=\pm i\omega$,  
corresponding to the collective mode 
and two damped modes,
\begin{align}\label{eq:DampedModes}
\Omega_{-,\pm}
= -\Gamma_{\text{int}}
\pm \sqrt{\delta \Gamma^{2} -\omega^{2} } 
= -\Gamma_{\text{int}}
\pm i\omega
+\mathcal{O}\left(
\frac{\delta\Gamma^2}{\omega^2}
\right),
\end{align}
corresponding to the relative mode. Here $\Gint$ was defined below \cref{eqRelativeMode} and $\delta\Gamma=\frac{1}{2\bar{w}^{\M} } \left( \frac{\varkappa^{\M}_{\T} }{3} -\varkappa^{\M}_{\V} \right) $.
In the second equality of \cref{eq:DampedModes}, we assume $\frac{\varkappa^{\M}_{\V/\T} }{\omega\bar{w}^{\M}}\ll 1$ and neglect the subleading terms within this expansion. 
The same approximation will be assumed for the mode coefficients of $v_\alpha^j(\tau,\mathbf{k})$ given in terms of initial conditions below. 
We will see that the quantity $\Gint$ is then the only dimensionful parameter describing the effect of the interactions between the dark and visible fluids on the velocity fields; consequently, we identify it with the interaction rate discussed in \cref{sec:DarkSectors}. 
In the regime~\eqref{eq:Timescales} considered in this work, the assumption $\frac{\varkappa^{\M}_{\V/\T} }{\omega\bar{w}^{\M}} \sim \Gint/\omega \ll 1$ is satisfied for the values of $\omega$ close to the peak of the kinetic spectra that have the largest influence on the GW spectrum. Indeed, in~\ref{sec:appNLO} we examine the leading corrections to this approximation and we observe that they do not affect the GW spectra in a visible way. Consequently, in the main text we work to leading order in the $\frac{\varkappa^{\M}_{\V/\T} }{\omega\bar{w}^{\M}}\ll 1$ expansion. 

The velocity fields obtained as solutions of the wave equations above are longitudinal, $v_\alpha^j(\tau,\mathbf{k}) = \hat{k}^j v_\alpha(\tau,\mathbf{k})$ with $\hat{k}^j\equiv k^j/k$, and 
\begin{equation}\label{eq:velocity_LO}
v_\alpha(\tau,\mathbf{k})= \frac{\bar{w}_d}{\bar{w}_d+\bar{w}_v} \sum_{\sigma = \pm} A_\sigma(\mathbf{k}) e^{i\sigma \omega (\tau-\tau_\ast)} \left(1+ R_\alpha e^{-\Gint (\tau-\tini)}\right)\,.
\end{equation}
Here $\tau_\ast$ is the initial time of sound wave propagation and $A_\sigma(\mathbf{k})$ is determined by the initial conditions in the dark fluid 
\begin{equation}\label{eq:ModeCoefficients_LO}
    A_\pm (\mathbf{k}) = \frac{1}{2} \left[v_d(\tau_\ast,\mathbf{k}) \pm c_s \lambda_d(\tau_\ast, \mathbf{k})\right],
\end{equation}
assuming vanishing initial perturbations in the visible fluid. The solutions for the dark and visible fluids differ only by the factor $R_\alpha$ determining the amplitude of the damped mode:
\begin{equation}\label{eq:Rfactors}
    R_d = \frac{\bar{w}_v}{\bar{w}_d}, \qquad R_v = -1.
\end{equation}
Consequently, the late-time evolution of the velocity fields is governed entirely by the collective mode $\Phi_+$, described by \cref{eqCollectiveMode}, and is therefore independent of the precise form of the right-hand side of the wave equations \eqref{wave_equations_eq6}. We stress that the solution for the collective mode in \cref{eq:velocity_LO} is exact, so the late-time limit is unaffected by the assumption $\frac{\varkappa^{\M}_{\V/\T}}{\omega\bar{w}^{\M}}\ll 1$.

\section{Gravitational-wave spectrum within the sound-shell model}\label{sec:GW}

\subsection{Tensor-mode perturbations sourcing GW}
The velocity fields \cref{eq:velocity_LO} give rise to anisotropic stresses which, in turn, source GWs.  
In this section, we describe the connection between the GW spectrum and the unequal-time correlator (UETC) of the anisotropic stresses, the latter will be then calculated in the following subsection. Similar discussion can be also found in~\cite{RoperPol:2023dzg}.

Let us consider the transverse-traceless (TT) tensor perturbations $l_{ij}$ in an expanding Universe, i.e. 
\begin{align}\label{eqMetricWithPerturbation}
    ds^{2}=a^{2}\left( \tau \right)  \left[ -d\tau^{2} +\left( \delta_{ij} +l_{ij}\right)  dx^{i}dx^{j}\right].
\end{align}
Perturbations $l_{ij}$ fulfill the linearized Einstein equation \cite{Baumann:2022mni,Caprini:2018mtu} in Fourier space
\begin{align}\label{eqLinEinstein1}
\left( \frac{d^{2}}{d\tau^{2} } +2\mathcal{H} \frac{d}{d\tau } +k^{2}\right)  l_{ij}\left( \tau ,\mathbf{k} \right)  =16\pi  Ga^{2}\bar{\rho}\Pi_{ij} \left( \tau ,\mathbf{k} \right),  
\end{align}
where $\mathcal{H}=a'/a=aH$ is the conformal Hubble factor  and $\Pi_{ij}$ is the dimensionless anisotropic stress defined as the TT-part of the perturbation of the energy-momentum tensor:
\begin{align}\label{eqDefOfAnisotropicStress}
    a^{2}\bar{\rho}\Pi_{ij}(\tau,\mathbf{k}) =\Lambda_{ijkl}(\hat{\mathbf{k}})\left( \delta T_{kl} (\tau,\mathbf{k})-a^2\bar{p}l_{kl}(\tau,\mathbf{k})\right).
\end{align}
Here $\bar{\rho }(\tau) =3\mathcal{H}^{2} /8\pi Ga^{2}$ is the total energy density and the TT-projector  reads
\begin{align}\label{eqTTProjector}
2\Lambda_{ijlm} &\equiv P_{il}P_{jm}+P_{im}P_{jl}-P_{ij}P_{lm}, \\
    P_{ij}\left( \hat{\mathbf{k} } \right) & =\delta_{ij} -\hat{k}_{i} \hat{k}_{j}.
\end{align}
For small acoustic perturbations discussed in Section \ref{secHydrodynamics},
\begin{align}\label{eq:TklFromVelocities}
a^{2}\bar{\rho } \Pi_{ij} (\tau ,\mathbf{k} )=\Lambda_{ijkl} (\hat{\mathbf{k} } )\sum^{}_{\alpha =v,d} \int \frac{\dd\mathbf{q} }{\left( 2\pi \right)^{3}  } a^{2}\bar{w}_{\alpha } v_{{\alpha ,k}  }\left( \tau,\mathbf{q} \right)  v_{\alpha ,l}\left( \tau,\mathbf{k}-\mathbf{q}\right)  
\end{align}
and we see that $\Pi_{ij} \left( \tau ,\mathbf{k} \right)$ is invariant under the transformations \eqref{eqMinkowskiRescaling}. 
For the radiation domination epoch, i.e. $a=\tau/\tau_\ast$ and $\mathcal{H}=1/\tau$, we get for the rescaled perturbations $h_{ij}=al_{ij}$
\begin{align}\label{eqLinEinstein2}
   \left( \frac{d^{2}}{d\tau^{2} } +k^{2}\right)  h_{ij}\left( \tau ,\mathbf{k} \right)  =\frac{6\mathcal{H}_{\ast } }{\tau } \Pi_{ij} \left( \tau ,\mathbf{k} \right).
\end{align}
If the source is active during $\tau \in \left[ \tau_{\ast } ,\tau_{\text{fin} } \right]$, for vanishing initial metric perturbations, one obtains:
\begin{align}\label{eqGreensFunction}
h_{ij}\left( \tau >\tau_{\text{fin} } ,\mathbf{k} \right)  =\frac{6\mathcal{H}_{\ast } }{k} \int^{\tau_{\text{fin} } }_{\tau_{\ast } } d\tau_{1} \frac{\Pi_{ij} \left( \tau_{1} ,\mathbf{k} \right)  }{\tau_{1} } \sin \left( k\left( \tau -\tau_{1} \right)  \right).  
\end{align}
The fractional energy density of GW today reads
\begin{align}\label{eqGWenergyDensity}
  &  \Omega_{\text{GW} } \left( \tau_{0} \right)  =\int^{\infty }_{-\infty } \Omega_{\text{GW} } \left( \tau_{0} ,k\right)  \text{d} \  \text{ln} k \nonumber \\
     &\equiv \frac{1}{32\pi G\bar{\rho}_{0}  } \left< \left( \dot{l}_{ij} \left( \tau_{0} ,\mathbf{x} \right)  \right)^{2}  \right>\approx \frac{1}{12\mathcal{H}^{2}_{0} a^{2}_{0}} \left< \left( h^{\prime }_{ij}\left( \tau_{0} ,\mathbf{x} \right)  \right)^{2}  \right>,
\end{align}
where in the last line terms proportional to $1/k\tau_0\ll1$ were neglected and the angular brackets denote an ensemble average over stochastic GW realizations.

The UETC spectrum of the anisotropic stresses is defined as
\begin{align}\label{eqEPiDefinition}
    \left< \Pi_{ij} \left( \tau_{1} ,\mathbf{k} \right)  \Pi^{\ast }_{ij} \left( \tau_{2} ,\mathbf{k}_{2} \right)  \right>  \equiv \left( 2\pi \right)^{6}  \delta^{3} \left( \mathbf{k} -\mathbf{k}_{2} \right)  \frac{\EPi \left( \tau_{1} ,\tau_{2} ,k\right)  }{4\pi k^{2}}. 
\end{align}
Then, using \cref{eqGreensFunction,eqGWenergyDensity,eqEPiDefinition} we get the GW spectrum today in terms of $\EPi$:
\begin{align}\label{eqGWinTermsOfEpi}
    \Omega_{\text{GW} } \left( \tau_{0} ,k\right)  &=3k\mathcal{T}_{\text{GW} } \int^{\tau_{\text{fin} } }_{\tau_{\ast } } \frac{d\tau_{1} }{\tau_{1} } \int^{\tau_{\text{fin} } }_{\tau_{\ast } } \frac{d\tau_{2} }{\tau_{2} }  \\&\times \EPi \left( \tau_{1} ,\tau_{2} ,k\right)  \cos k\left( \tau_{0} -\tau_{1} \right)  \cos k\left( \tau_{0} -\tau_{2} \right),\nonumber  
\end{align}
where the transfer function $\mathcal{T}_{\text{GW} }$ reads 
\begin{align}\label{eqTransferFunction}
    h^{2}\mathcal{T}_{\text{GW} } \equiv \left( \frac{a_{\ast }}{a_{0}} \right)^{4}  \left( \frac{H_{\ast }}{H_{0}/h} \right)^{2}  \approx 1.6\times 10^{-5}\left( \frac{100}{g_{\tot,\sym }} \right)^{1/3}, 
\end{align}
with the total number of relativistic degrees of freedom $g_{\tot,\sym }$ defined below \cref{eq:alphatot}.
\cref{eqTransferFunction} is derived assuming the entropy conservation in the thermal bath and also using $\Hast = H_\ast$ that holds for the choice $a_\ast = 1$. Expanding the product of the Green’s functions with the product-to-sum identities and averaging over the rapidly oscillating terms with $k\tau_0 \gg 1$, we obtain
\begin{align}\label{eqTransferFunctionFinal}
    \Omega_{\text{GW} } \left( \tau_{0} ,k\right)  &\approx\frac{3k}{2}\mathcal{T}_{\text{GW} } \int^{\tau_{\text{fin} } }_{\tau_{\ast } } \frac{d\tau_{1} }{\tau_{1} } \int^{\tau_{\text{fin} } }_{\tau_{\ast } } \frac{d\tau_{2} }{\tau_{2} } \nonumber \\&\times \EPi \left( \tau_{1} ,\tau_{2} ,k\right)  \cos k\left( \tau_{1} -\tau_{2} \right). 
\end{align}

\subsection{Anisotropic stresses from sound waves}

In this section, we present the derivation of the UETC spectrum of anisotropic stresses $E_\Pi$, 
including contributions of the dark and visible fluids. We generalize the one-fluid calculation within the sound-shell model~\cite{Hindmarsh:2016lnk,Hindmarsh:2019phv,RoperPol:2023dzg}, following mainly the derivation from~\cite{RoperPol:2023dzg}.

Formula~\eqref{eq:TklFromVelocities} implies that the two-point correlator of the anisotropic stresses~\eqref{eqEPiDefinition} includes four-point correlators of the velocity fields. Assuming that these stochastic fields are Gaussian as in~\cite{Hindmarsh:2019phv,RoperPol:2023dzg}, Wick's theorem can be used to obtain:
\begin{align}\nonumber
    \bar{\rho}^2 \langle \Pi_{ij}(\tau_1,\mathbf{k}) \Pi_{lm} (\tau_2,\mathbf{k_2})\rangle =\;  \sum_{\alpha=d,v} \sum_{\alpha'=d,v} \bar{w}_\alpha \bar{w}_{\alpha'} \int \frac{\dd^3\mathbf{p}}{(2\pi)^3} \int \frac{\dd^3\mathbf{q}}{(2\pi)^3} \\ \nonumber
    \times\, \biggl[
    \langle v_{\alpha,i}(\tau_1, \mathbf{p})v_{\alpha'\!,l}^\ast (\tau_2, \mathbf{q})\rangle \langle v_{\alpha,j}(\tau_1, \mathbf{k-p})v_{\alpha'\!,m}^\ast (\tau_2, \mathbf{k_2-q})\rangle  \\ 
    \label{eq:corTijWick2Fluid}
   + \langle v_{\alpha,i}(\tau_1, \mathbf{p})v_{\alpha'\!,m}^\ast (\tau_2, \mathbf{k_2-q})\rangle \langle v_{\alpha,j}(\tau_1, \mathbf{k-p})v_{\alpha'\!,l}^\ast (\tau_2, \mathbf{q})\rangle  \biggr]\,
\end{align}
The velocity fields induced by the sound wave propagation are irrotational, hence, in analogy with notation used in~\cite{RoperPol:2023dzg}, the two-point correlation functions can be written as:
\begin{equation}\label{eq:defEkinTwoFluid}
    \langle v_{\alpha,i}(\tau_1, \mathbf{k})v_{\alpha'\!,j}^\ast (\tau_2, \mathbf{k_2})\rangle = (2\pi)^6 \hat{k}_i \hat{k}_j \delta^3(\mathbf{k}-\mathbf{k_2})\frac{2E_{\kin}^{\alpha\alpha'}(\tau_1,\tau_2,k)}{4\pi k^2}\,.
\end{equation}
Let us note that an additional term would appear in~\eqref{eq:corTijWick2Fluid}, however, this term vanishes after inserting~\eqref{eq:defEkinTwoFluid} and applying the transverse traceless projector. Using similar manipulations as in~\cite{RoperPol:2023dzg}, we arrive at
\begin{align}\nonumber
    E_\Pi(\tau_1,\tau_2,k) =  \frac{2k^2}{\bar{\rho}^2} \sum_{\alpha=d,v} \sum_{\alpha'=d,v} \bar{w}_\alpha \bar{w}_{\alpha'} \int_{-1}^1 \dd z \int_0^\infty \dd p \\ \label{eq:EPiTwoFluid}
  \times \, \frac{p^2}{\tilde{p}^4}\,(1-z^2)^2\,  E_\kin^{\alpha,\alpha'} (\tau_1,\tau_2,p)\,E_\kin^{\alpha,\alpha'} (\tau_1,\tau_2,\tilde{p})\,,
\end{align}
where $\tilde{\mathbf{p}} = \mathbf{k} - \mathbf{p}$ and $z=\hat{\mathbf{k}} \cdot \hat{\mathbf{p}}$. 

The key task is to evaluate $E_\kin^{\alpha,\alpha'} (\tau_1,\tau_2,k)$ based on the solutions of the wave equations~\cref{eq:velocity_LO}. Within the sound-shell model~\cite{Hindmarsh:2019phv,RoperPol:2023dzg}, the initial conditions for the velocity and energy density fields are based on the fluid profiles of a set of $N_b$ bubbles\footnote{In the context of our work, we apply this initial condition to the dark fluid and assume a zero initial condition in the visible fluid. The assumption of small temperature and velocity differences between the fluids used in \cref{eq:2_fluid_current} may not be fully valid then. On the other hand, even in the original sound-shell model, similar assumption is made, since the linearized equations are used straight after the phase transition completion. Consequently, we still work with the near-equilibrium expansion here, but it would be interesting to compare our results with full hydrodynamic simulations in the future.} nucleated at different times $\tau_0^{(n)}$ and positions $\mathbf{x}_0^{(n)}$. Denoting the n$^{\text{th}}$ bubble lifetime by $T_n = \tau_\ast - \tau_0^{(n)}$ and introducing a dimensionless variable $\chi = k T_n$, the amplitudes~\eqref{eq:ModeCoefficients_LO} can be written as
\begin{equation}\label{eq:AsumBubbles}
    A_\pm(\mathbf{k}) = \sum_{n=1}^{N_b} \A_{\pm}(\chi) T_n^3 e^{i\mathbf{k}\cdot \mathbf{x}_0^{(n)}}\,
\end{equation}
where 
\begin{equation}\label{eq:CalADeffl}
    \A_\pm (\chi) = -\frac{i}{2} \left[f'(\chi) \pm i c_s l(\chi)\right]\, ,
\end{equation}
$f'(\chi) = \dd f(\chi)/\dd\chi$,
and
\begin{align}\label{eq:fchilchi}
    f(\chi) &=\! \frac{4\pi}{\chi} \!\int_0^\infty \! \dd\xi v_\ip(\xi) \sin (\chi\xi)\, , \\
    l(\chi) &=  \!\frac{4\pi}{\chi}\! \int_0^\infty \! \dd\xi \xi \lambda_\ip(\xi) \sin (\chi\xi)\, .
\end{align}
Here $v_\ip$ and $\lambda_\ip$ are the single-bubble radial profiles that depend on the phase transition parameter $\alpha_d$ from \cref{eq:alphad} and the bubble wall velocity~$v_w$, and we calculate them following Refs.~\cite{Espinosa:2010hh,Hindmarsh:2016lnk,Caprini:2026nnk}.\footnote{Let us note that the energy density profiles of the bubbles are defined as $\lambda_\ip(\xi) \equiv (\rho(\xi)-\rho_\mathrm{sym})/w_n$ where $\rho_\mathrm{sym}$ is the energy density in the symmetric phase and $w_n$ is the dark sector enthalpy density at nucleation~\cite{Caprini:2026nnk}. In order to match $\lambda_\ip(\xi)$ with the initial condition for $\lambda_d$, $\lambda_\ip$ should be, hence, in principle rescaled by $w_n/\bar{w}_d$ and shifted by a constant. Ref.~\cite{Caprini:2026nnk} shows that the mean enthalpy differs from $w_n$ by at most a few percent, consequently, we put $w_n/\bar{w}_d \approx 1$. The constant shift, in turn, contributes only to the zero momentum mode of $\lambda_d$ which is irrelevant for the calculation of the GW spectrum.} In terms of the rescaled radial coordinate $\xi = R_n/T_n$ with $R_n = \mathbf{x} - \mathbf{x}_0^{(n)}$, the initial bubble profiles are universal, hence, $\A_\pm (\chi)$ depends on $n$ only through $\chi = T_n k$. In the velocity correlator~\eqref{eq:defEkinTwoFluid}, average over the $N_b$ randomly nucleated bubbles in a volume $\mathcal{V}$ can then be performed~\cite{Hindmarsh:2016lnk}. In the infinite volume limit, $N_b/\mathcal{V} \to R_\ast^{-3}$ 
is obtained and the sum over bubbles can be replaced by an integral over the bubble lifetimes, leading to 
\begin{align}\nonumber
    & E_\kin^{\alpha,\alpha'} (\tau_1, \tau_2, k) = F_\alpha (\tau_1)F_{\alpha'} (\tau_2)\sum_{\sigma = \pm} \sum_{\sigma'=\pm} e^{i\omega\left[\sigma(\tau_1-\tau_\ast) - \sigma'(\tau_2 - \tau_\ast)\right]} \\ \label{eq:EkinSumSigns}
    & \times \frac{k^2}{4\pi^2 R_\ast^3 \beta^6} \int \dd\tilde{T} \tilde{T}^6 \nu(\tilde{T}) \A_{\sigma} (k \tilde{T}/\beta) \A_{\sigma'}^\ast(k \tilde{T}/\beta)\,.
\end{align}
Here 
\begin{equation}
F_\alpha = \frac{\bar w_d}{\bar w_d + \bar w_v} \left(1+ R_\alpha e^{-\Gint (\tau-\tini)}\right), 
\end{equation}
$\tilde{T}$ is the bubble lifetime rescaled by the inverse phase transition duration $\beta$, and $\nu$ denotes the distribution function for $\tilde{T}$.\footnote{It was shown in \cite{Hindmarsh:2019phv} that the absolute time of bubble nucleation and collision drops out from the kinetic spectrum calculation, the kinetic spectrum depends only on the distribution of the bubble lifetimes. Consequently, the results for the GW spectrum should depend only on $\tau_\sw = \tau_\fin - \tau_\ast$, not on $\tau_\ast$ itself, see~\cite{Caprini:2026nnk} for further discussion of this assumption. In our case, however, $\tau_\ast$ determines also the starting point of the energy and momentum transfer between the dark and visible fluids. This is, of course, somewhat artificial, the correct approach would be to include the coupling between the fluids already at the stage of bubble growth. This correction is, however, left for future work.} For easier comparison with the main results of~\cite{RoperPol:2023dzg}, we assume the scenario of exponential nucleation which gives rise to $\nu(\tilde{T}) = e^{-\tilde{T}}$. For the difference with the choice of simultaneous nucleation, see~\cite{Hindmarsh:2019phv,RoperPol:2023dzg}.

It turns out that \cref{eq:EkinSumSigns} can be related to the single-fluid analog $E_\kin$ calculated in~\cite{RoperPol:2023dzg} by a simple relation $E_\kin^{\alpha,\alpha'} (\tau_1, \tau_2, k) = F_\alpha (\tau_1)F_{\alpha'} (\tau_2) E_\kin (\tau_1, \tau_2, k)$. Consequently, very similar steps as in~\cite{RoperPol:2023dzg} can be taken. As in~\cite{Hindmarsh:2019phv,RoperPol:2023dzg} we neglect the terms containing $\sin$ or $\cos$ functions with arguments proportional to $\tau_1+\tau_2-2\tau_\ast$ since these contributions are highly oscillatory for the momenta around the peak of the kinetic spectrum and can be neglected after integration over momenta (although they become important in the $k\to 0$ limit as argued in~\cite{RoperPol:2023dzg}). Consequently,
\begin{equation}\label{eq:EkinFin}
   E_\kin^{\alpha,\alpha'} (\tau_1, \tau_2, k) \approx  F_\alpha (\tau_1)F_{\alpha'} (\tau_2) E_\kin(k) \cos \omega(\tau_1-\tau_2)
\end{equation}
where
\begin{align}\label{eq:Ekink}
&E_\kin (k) =  \frac{k^2}{2\pi^2 R_\ast^3 \beta^6} \int \dd\tilde{T} \tilde{T}^6 \nu(\tilde{T}) \E (k \tilde{T}/\beta)\, ,\\
&\E (\chi) = |\A_+(\chi)|^2 = \frac{1}{4}\left[f'(\chi)^2 + c_s^2 l(\chi)^2\right]\,.
\end{align}
In order to evaluate the UETC spectrum of the anisotropic stresses~\eqref{eq:EPiTwoFluid}, we define:
\begin{equation}\label{eq:sumAlpha}
    \frac{1}{\bar{w}_d^2}\,\sum_{\alpha=d,v} \sum_{\alpha'=d,v} \bar{w}_\alpha \bar{w}_{\alpha'} [F_\alpha(\tau_1)]^2 [F_{\alpha'}(\tau_2)]^2 \equiv  F(\tau_1) \, F(\tau_2)
\end{equation}
with
\begin{equation}\label{eq:defFtot}
    F(\tau) = \frac{1 + \frac{\bar{w}_v}{\bar{w}_d}\, e^{-2\Gint(\tau - \tau_\ast)}}{1 + \frac{\bar{w}_v}{\bar{w}_d}}\,.
\end{equation}

\subsection{GW spectrum}
Combining formulas~\eqref{eqTransferFunctionFinal},~\eqref{eq:EPiTwoFluid},~\eqref{eq:EkinFin} and~\eqref{eq:sumAlpha}, the GW spectrum can be written as
\begin{align}\nonumber
 \Omega_\text{GW}(\deltaufin,k) &= 
    3 \, \frac{\bar w_d^2}{\bar{\rho}^2}  \, k^3 \, {\cal T}_{\rm GW} \int_{-1}^1 \bigl(1 - z^2 \bigr)^2 \, \dd z 
    \int_0^\infty \!\! \dd p \,\\ \label{eq:omGWfinal}
    &\times \frac{p^2}{\tilde p^4} \, E_\kin (p)
    E_\kin (\tilde p) \, 
    \Delta (\deltaufin, k, p, \tilde p) \, 
\end{align}
where
\begin{align}\nonumber
   \Delta (\deltaufin,  k, p, \tilde p) \equiv 
    \int_{\tini^{ }}^{\tfin^{ }} \,\frac{\dd \tau^{ }_1}{\tau^{ }_1} 
    \int_{\tini^{ }}^{\tfin^{ }} \, \frac{\dd \tau^{ }_2}{\tau^{ }_2} \, \\ \label{eq:Delta_2f}
    \times \cos (p c_s \tau^{ }_-)
    \cos (\tilde p c_s \tau^{ }_-)
    \cos (k\tau^{ }_-)  F(\tau_1) \, F(\tau_2)\,
\end{align}
with $\tau_- \equiv \tau_1-\tau_2$ and $\deltaufin \equiv \tau_\fin - \tini$. Further, $\bar{\rho}$ is the total energy density that drives the Universe expansion, i.e., it can be identified with $\bar{\rho}_{\tot,\sym}$ appearing in \eqref{eq:alphatot}. Using the bag equation of state, one gets
\begin{equation}\label{eq:facwrho}
    \frac{\bar w_d^2}{\bar{\rho}^2} \equiv \frac{\bar w_d^2}{\bar{\rho}_{\tot,\sym}^2} = \frac{\bar{w}_d^2}{(\bar{\rho}_{\tot,\sym}-\epsilon)^2(1+\alpha_\tot)^2} = \frac{4}{3} \frac{g_{d,\bro}^2}{g_{\tot,\sym}^2}\frac{1}{(1+\alpha_\tot)^2}\,.
\end{equation}
The factor $g_{d,\bro}^2/g_{\tot,\sym}^2$ gives overall suppression of the GW spectrum compared to the phase transitions occurring in the Standard Model plasma, both for secluded dark sector and weakly coupled dark sector.\footnote{Let us note that in our numerical results, we evaluate the factor from \cref{eq:facwrho} assuming $g_{d,\bro} = g_{d,\sym}$ which seems to be reasonably well satisfied for weak phase transitions~\cite{Biondini:2026uds}. Change in this factor would lead just to an overall shift in amplitude of all the GW spectra presented in \cref{sec:numerics}.}

The transfer of energy and momentum between the dark and visible sectors is, in turn, captured by the factors $F(\tau_1)F(\tau_2)$ defined in~\cref{eq:defFtot}. Let us note that $F(\tau) \to 1$ for $\Gint \to 0$, and this limit corresponds to a secluded dark sector. On the other hand, $F(\tau) \to \bar{w}_d /(\bar{w}_d + \bar{w}_v) < 1$ for large values of $\Gint (\tau-\tau_\ast)$, hence, if $\Gint \tau_{\sw}$ is non-negligible, suppression of the GW spectrum compared to the secluded dark sector case is expected. This suppression is a direct consequence of the suppression of the velocity field amplitudes, see~\cref{eq:velocity_LO}. Indeed, while in the secluded dark sector case the anisotropic stress $\bar{\rho}\Pi_{ij}$  
from \cref{eq:TklFromVelocities} is schematically proportional to $\bar{w}_d v_{d,\mathrm{sec}}^2$, this quantity becomes proportional to $\sum_{\alpha = v,d} \bar{w}_\alpha \bar w_d^2/(\bar w_d + \bar w_v)^2 v_{d,\mathrm{sec}}^2 = w_d^2/(\bar w_d + \bar w_v)\, v_{d,\mathrm{sec}}^2$ for large $\Gint (\tau-\tau_\ast)$ which explains the suppression of the resulting GW signal.

The function $\Delta(\deltaufin, k, p, \tilde p)$ from~\cref{eq:Delta_2f} is in general a highly oscillating function that can be rewritten in terms of exponential integrals $\mathrm{Ei}$, see the accompanying Mathematica code~\cite{git_SSModel}.
However, for certain ranges of momenta, it can be approximated by simpler expressions. First, for the points where $\min (k,p) \times c_s \deltaufin \gg 1$, this function 
acts as:
\begin{equation}\label{eq:DeltaHighk}
    {\Delta} (\deltaufin, k, p, \tilde p) \approx \frac{\pi}{2} \, \delta\left(c_s(p+\tilde{p})-k\right) \int_{\tau_{\ast}}^{\tau_\fin} \frac{\mathrm{d}\tau}{\tau^2}F(\tau)^2\,.
\end{equation}
The function $E_\kin(p)$ peaks around $p R_\ast \sim 1$, which induces the peak of the GW spectrum at $k R_\ast \sim 1$. Consequently, for the benchmark values $H_\ast R_\ast \lesssim 0.01$ and $\deltaufin H_\ast \gtrsim 0.1$, the approximation~\eqref{eq:DeltaHighk} is expected to be valid around the peak values of the GW spectrum. This allows us to find the suppression of the peak GW amplitude for the weakly coupled dark sectors compared to the secluded dark sector:
\begin{equation}\label{eq:Upsilon}
    \frac{\Omega_{\GW,\,\mathrm{peak}}}{\Omega_{\GW,\,\mathrm{peak}}^{\Gint \to 0}} = \frac{\Upsilon}{\Upsilon(\Gint\to 0)}\,,\quad \Upsilon = \frac{1}{H_\ast}\int_{\tau_{\ast}}^{\tau_\fin} \frac{\mathrm{d}\tau}{\tau^2}F(\tau)^2\,.
\end{equation}
Using $H_\ast \tini = 1$, we obtain $\Upsilon(\Gint\to 0) = H_\ast \deltaufin / (1+H_\ast \deltaufin)$, which recovers the dependence of the peak amplitude on the source duration observed in~\cite{RoperPol:2023dzg}.

In the opposite limit, $k\to 0$, the function $\Delta$ can be approximated for the values of $p$ around the peak of the kinetic spectrum as
\begin{equation}\label{eq:DeltaLowk}
    \Delta (\deltaufin, k, p, \tilde p) \xrightarrow{k\to 0}   \Delta_0 \equiv \frac{1}{2} \left(\int_{\tini}^{\tau_\fin} \frac{\dd \tau}{\tau} F(\tau) \right)^2\,
\end{equation}
provided $c_s\deltaufin/R_\ast \gg 1$~\cite{RoperPol:2023dzg}. Consequently, the low-frequency tail is suppressed compared to the case of the secluded dark sector by the $k$-independent factor $\Delta_0/\Delta_0(\Gint \to 0)$ where $\Delta_0(\Gint \to 0) = 1/2 \ln^2(1+\deltaufin H_\ast)$ in agreement with the results of~\cite{RoperPol:2023dzg}. Importantly, this implies that the GW spectrum grows as $k^3$ at low $k$.

\section{Numerical results}\label{sec:numerics}
\begin{figure*}[ht!]
    \centering
    \includegraphics[width=\textwidth]{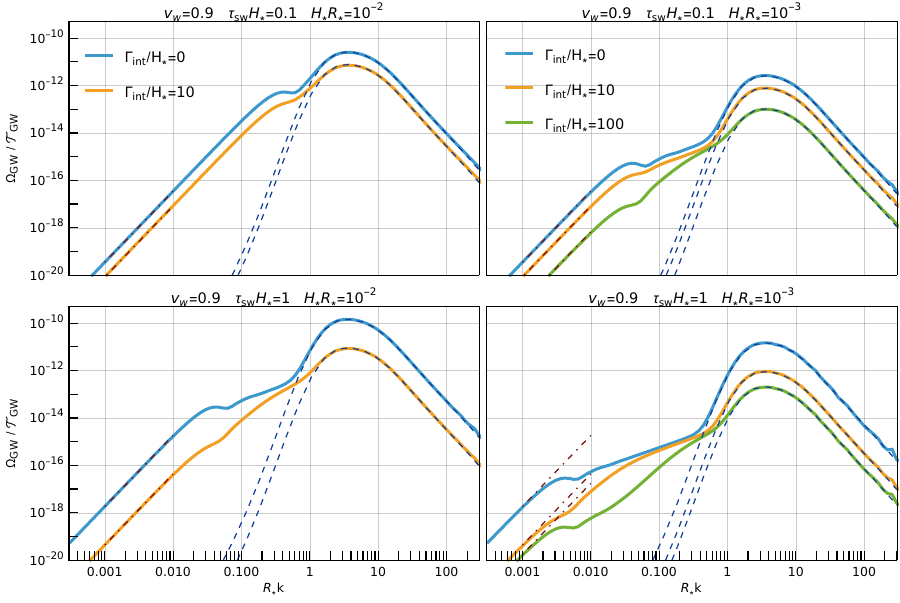}
    \caption{
    Gravitational-wave spectra for a wall velocity $v_w=0.9$ and transition strength $\alpha=0.1$ computed from \cref{eq:omGWfinal}, up to the redshift factor $\mathcal{T}_\text{GW}$.
    The top and bottom panels correspond to source durations $\deltaufin=0.1$ and $1$, respectively, while the left and right panels correspond to $H_\ast R_\ast=10^{-2}$ and $10^{-3}$, respectively.
    The dark-red dot-dashed and dark-blue dashed lines show the low- and high-frequency approximations given in \cref{eq:DeltaLowk,eq:DeltaHighk}, respectively.
    Results obtained with an extended version of \texttt{cosmoGW} \cite{RoperPol:cosmoGW,git_SSModel}.
    }
    \label{fig:GW}
\end{figure*}

The integrals in~\cref{eq:omGWfinal} must be evaluated numerically, and we present the resulting GW spectra in \cref{fig:GW} for several benchmark parameter choices.
We release the \texttt{Mathematica} package \texttt{SSModel}~\cite{git_SSModel}, which implements both the standard sound-shell model and its extension to two sectors developed in this work.
The package includes the evaluation of 
the time integral in~\cref{eq:Delta_2f}, 
the GW spectrum in~\cref{eq:omGWfinal} and its approximations in~\cref{eq:DeltaLowk,eq:DeltaHighk}, the suppression factor in~\cref{eq:Upsilon}, and the NLO corrections discussed in~\labelcref{sec:appNLO}.
The GW spectra can be equivalently obtained using an extension of the \texttt{cosmoGW} code~\cite{RoperPol:cosmoGW}, implementing \cref{eq:omGWfinal,eq:Delta_2f}, which is also available in the 
repository~\cite{git_SSModel}. \cref{fig:GW} was produced in the latter way for computational efficiency.
We choose two benchmark values for the rescaled mean bubble separation: $R_\ast H_\ast = 10^{-2}$, and $10^{-3}$. The interaction rate between the dark and visible sectors is then chosen so that $\Gint/H_\ast$ is at least 10 times smaller than $1/(R_\ast H_\ast)$, in order to satisfy the condition~\eqref{eq:Timescales}.  For an easy comparison with~\cite{RoperPol:2023dzg}, we use the same benchmark for the phase transition strength: $\alpha_d = 0.1$. For such a moderate phase transition, the development of non-linearities is expected.  For a secluded dark sector, $\tau_{\mathrm{nl}}$ can be estimated based on the fraction $K$ of energy in the bulk fluid motion as discussed in \cref{sec:DarkSectors}. However, when the fluid motion is transferred from the dark fluid to the visible one, the non-linearities may in principle develop later. For this reason, we remain agnostic and discuss the results for several values of $\tau_{\sw}$. Similarly, the calculations of bubble wall velocity $v_w$ are still subject to uncertainties (see, e.g.,~\cite{vandeVis:2025plm}), hence, we simply choose a single benchmark value $v_w=0.9$ to demonstrate our results. For this value, the bubbles of the new phase expand in the detonation regime, where the GW spectra from hydrodynamic simulations and from the sound-shell model seem to match best~\cite{Hindmarsh:2017gnf,Hindmarsh:2019phv}. The last parameter determining the GW spectrum is the ratio of the enthalpies in the visible and dark sectors $\bar{w}_v/\bar{w}_d = g_{v}/g_{d,\bro}$, determining the factor \eqref{eq:defFtot}. As discussed in \cref{sec:DarkSectors}, this ratio can be large in realistic scenarios, hence, we choose $\bar{w}_v/\bar{w}_d = 10$ in \cref{fig:GW}. Let us note that we do not specify $\mathcal{T}_\GW$ and rescale the GW frequency by $R_\ast$, consequently, our plots are independent of the phase transition temperature. 

\begin{figure*}
    \centering
    \includegraphics[width=\linewidth]{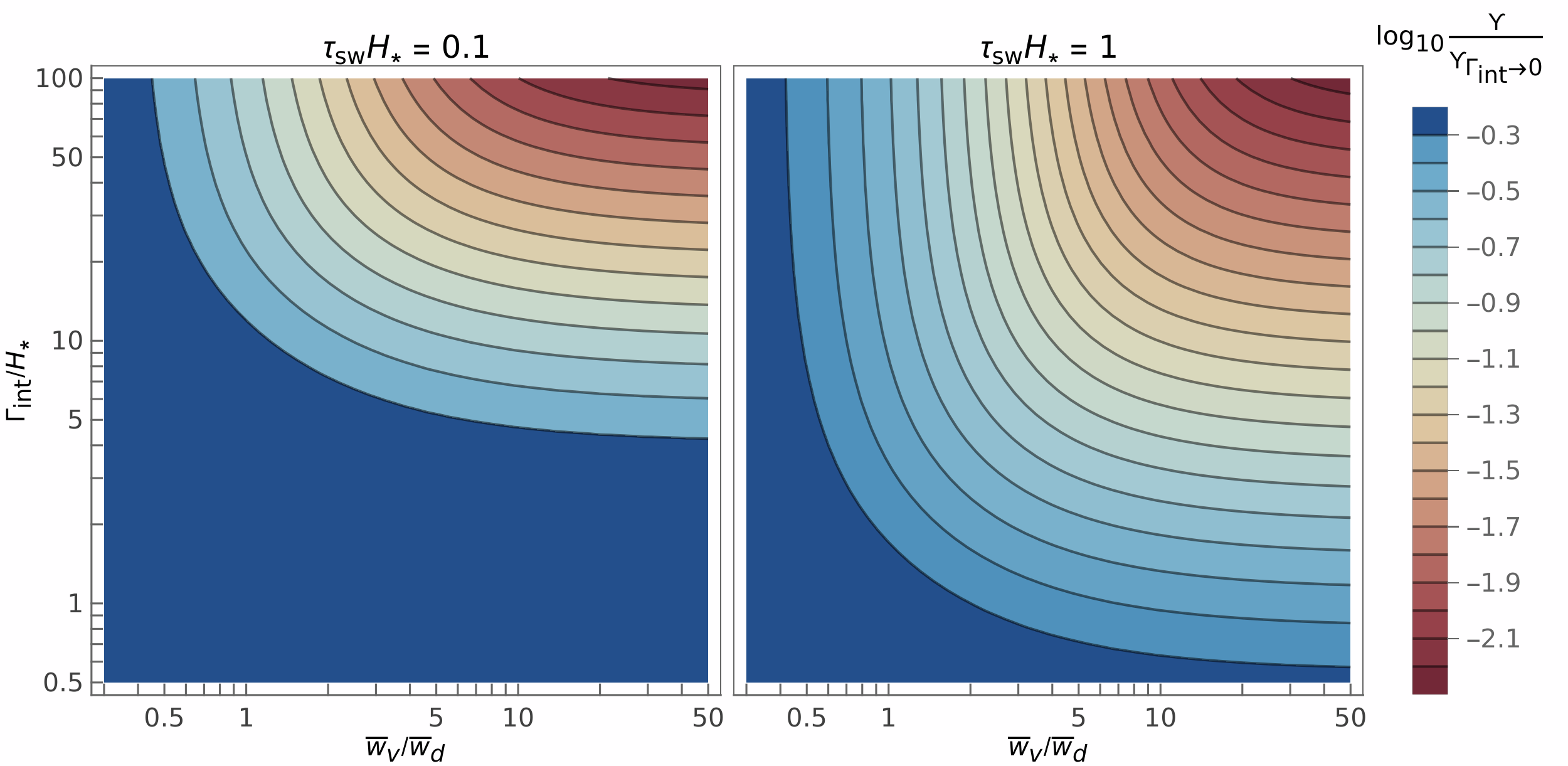}
    \caption{The factor $\Upsilon/\Upsilon_{\Gamma_\text{int}\rightarrow 0}$ from
    \cref{eq:Upsilon} describing the suppression of the peak amplitude of the GW spectrum compared to the secluded dark sector case. Left and right panels correspond to the sound wave source durations $\tau_\sw H_\ast =0.1$ and 1, respectively.
    }
    \label{fig:prefactor}
\end{figure*}

The dashed and dot-dashed lines in \cref{fig:GW} show the approximations given by formulas~\eqref{eq:DeltaHighk} and~\eqref{eq:DeltaLowk}, respectively. We observe that the former approximation describes well the spectrum at high frequencies and we checked that this holds also for other choices of $\bar{w}_v/\bar{w}_d$. Consequently, the factor~\eqref{eq:Upsilon} can be indeed used to approximate the suppression of the peak GW amplitude for the values of $H_\ast R_\ast$ and $\tau_\sw H_\ast$ used in \cref{fig:prefactor}, see \cref{fig:prefactor} for the value of this factor~\eqref{eq:Upsilon} for a range of $\Gint$ and $\bar{w}_v/\bar{w}_d$. As expected, the peak suppression is more pronounced for longer duration of sound wave propagation. 
Let us note that for $\tau_\sw H_\ast = 0.01$ the GW spectrum is suppressed even less; e.g., for $\Gint/H_\ast = 100$ and $\bar{w}_v/\bar{w}_d = 10$, one obtains $\Upsilon/\Upsilon_{\Gint \to 0} \sim 0.3$, while for $\tau_\sw H_\ast = 0.1$, this factor reads $\sim 0.04$. Let us also note that the approximation~\eqref{eq:DeltaHighk} breaks down around the peak of the spectrum for $R_\ast H_\ast \gtrsim 10^{-2}$ and $\tau_\sw H_\ast \lesssim 0.01$. 

In the high- and low-frequency ranges where the approximations~\eqref{eq:DeltaHighk} and~\eqref{eq:DeltaLowk} are valid, the shapes of the GW spectra coincide with the case of the secluded dark sector, i.e., they can be described by the fits from~\cite{RoperPol:2023dzg} up to an overall normalization. In the intermediate frequency region, however, the shape of the GW spectrum is modified which is one of our important results. Of course, the shape of the GW spectrum strongly depends on the bubble wall velocity and we plan to perform a more detailed study of the modifications due to $\Gint \neq 0$ in a follow-up work.

\section{Summary and outlook} \label{sec:conclusions}
In this work, we present modifications of the GW signal coming from a phase transition in a dark sector when this sector is weakly coupled to the SM plasma.\footnote{
    We note that the results presented in this paper apply equally to phase transitions in the visible sector, in the presence of a dark sector weakly coupled to the SM plasma. In this case, the roles of $\bar{w}_d$ and $\bar{w}_v$ are interchanged.
    }
We focus on the regime~\eqref{eq:Timescales}, where the fluid coupling can be neglected until the phase transition completes, but affects the subsequent, longer era of sound wave propagation. This era can be described by linearized equations, which we solve analytically to predict the GW signal within the sound-shell model~\cite{Hindmarsh:2016lnk,Hindmarsh:2019phv,RoperPol:2023dzg}. We find that the GW spectrum is generically suppressed compared to the secluded dark sector case, while its shape is also modified, see \cref{fig:GW}. Although these qualitative results are expected to be robust, several quantitative refinements remain possible.

First, the current framework can be generalized in several ways. For example, the sound speed in the broken phase is expected to deviate from $c_s^2 = 1/3$ for realistic dark sectors (see, e.g.,~\cite{Biondini:2026uds}) and such modification can be taken into account within the sound-shell model~\cite{Giombi:2024kju}. If the equations of state for the dark and visible sectors differ in the sound wave era, the discussion around \cref{eqCollectiveMode} will become more involved, likely implying a richer mode structure. Another task for future work is to include the coupling between the fluids in the derivation of the bubble profiles, similarly as in~\cite{Wada:2026iux}.

Second, the hydrodynamic description presented here needs to be supplemented by calculation of the phenomenological parameters $\varkappa_{\V/\T}$ describing the exchange of energy and momentum between the dark and visible fluids, see \cref{eq:2_fluid_current}. Given a specific microscopic model, these parameters can be obtained from collision terms in the Boltzmann equations and then used to calculate $\Gint$.

Finally, although the sound-shell model was shown to agree relatively well with the results of hydrodynamic simulations~\cite{Hindmarsh:2017gnf,Hindmarsh:2019phv} and was, consequently, used in different contexts for its simplicity~\cite{Guo:2020grp,Giombi:2024kju,Barni:2025gnm}, there are questions that cannot be answered within this framework. As already mentioned above, the timescale of the sound wave propagation $\tau_\sw$ needs to be determined from hydrodynamic simulations; in fact, the sound wave source may not be constant in time already for moderate phase transitions~\cite{Caprini:2024gyk}. Consequently, we believe that it would be interesting to study the effect of the coupling between the dark and visible fluids within the hydrodynamic simulations.

\section*{Acknowledgements}


The authors would like to thank Mikko Laine for insightful discussions and for valuable 
comments on the manuscript. They are also thankful to Simona Procacci, Alberto Roper Pol, Philipp Schicho, and Jorinde van de Vis for useful discussions. 

The work of H.K. and D.K. is supported by the Research Council of Norway under the FRIPRO Young Research Talent grant no. 335388. MF is supported by LIP and by the
Portuguese Foundation for Science and Technology (FCT), reference LA/P/0016/202, and from ERC-PT A-Projects ``Unveiling'', financed by PRR, NextGenerationEU.
MF is directly funded by FCT under the grant no.\ PRT/BD/154730/2023 (\url{https://doi.org/10.54499/PRT/BD/154730/2023}).
MF is grateful to the Department of Mathematics and Physics at the University of Stavanger for its hospitality throughout the development of this work and for its financial support of academic activities.

\appendix

\begin{strip}
\section{NLO analysis of the damped modes}\label{sec:appNLO}
As discussed below \cref{eq:DampedModes}, the results presented in the main text were obtained based on velocity fields given by formula~\eqref{eq:velocity_LO} where $\frac{\varkappa^{\M}_{\V/\T} }{\omega \bar{w}^{\M} } \ll 1$ is assumed. When including the leading corrections within this expansion, \cref{eq:velocity_LO} is replaced by:
\begin{equation}\label{eq:velocity_NLO}
v_\alpha(\tau,\mathbf{k})= \frac{\bar{w}_d}{\bar{w}_d+\bar{w}_v} \sum_{\sigma = \pm}  e^{i\sigma \omega (\tau-\tau_\ast)} \left\{A_\sigma(\mathbf{k}) + \left[A_\sigma(\mathbf{k}) + i \sigma \frac{\Gamma_V}{3\omega} v_d(\tini,\mathbf{k}) \right]R_\alpha e^{-(\Gint + i\sigma \frac{\delta \Gamma^2}{2\omega}) (\tau-\tini)}\right\}\,,
\end{equation}
where $\Gamma_V \equiv \varkappa^{\M}_{\V}/(2 \bar{w}a^{\M})$, $\delta\Gamma$ was introduced below~\eqref{eq:DampedModes}, and $A_\sigma(\mathbf{k})$ and $R_\alpha$ were defined in \cref{eq:ModeCoefficients_LO,eq:Rfactors}, respectively. 
As anticipated in the main text, the corrections affect the damped relative mode only. Both the amplitude and the frequency of this mode are affected, and we considered them separately in order to understand their qualitative effect. First, the correction to the frequency leads to the following modification of the function $\Delta$ from \cref{eq:Delta_2f}: 
\begin{align}\nonumber
   \Delta^{\LO+\NLO} (\deltaufin, k, p, \tilde p) = \frac{1}{4} \frac{\bar{w}_d^2}{(\bar{w}_d+\bar{w}_v)^2}
    \int_{\tini^{ }}^{\tfin^{ }} \,\frac{\dd \tau^{ }_1}{\tau^{ }_1} 
    \int_{\tini^{ }}^{\tfin^{ }} \, \frac{\dd \tau^{ }_2}{\tau^{ }_2} \,
\sum_{m,n \in \{-1,1\}} \biggl\{ \cos [(\tau_1-\tau_2) (k + m\omega + n\tilde{\omega})] \\ \nonumber
+ \frac{\bar{w}_v}{\bar{w}_d} e^{-2\Gint(\tau_1-\tini)} \cos \left[(\tau_1-\tau_2) (k + m\omega + n\tilde{\omega}) + (\tau_1-\tini) \left(m \omega \frac{\delta\Gamma^2}{2\omega^2} + n \tilde\omega\frac{\delta\Gamma^2}{2\tilde\omega^2}\right) \right]\\ \nonumber
+ \frac{\bar{w}_v}{\bar{w}_d} e^{-2\Gint(\tau_2-\tini)} \cos \left[(\tau_1-\tau_2) (k + m\omega + n\tilde{\omega}) - (\tau_2-\tini) \left(m \omega \frac{\delta\Gamma^2}{2\omega^2} + n \tilde\omega\frac{\delta\Gamma^2}{2\tilde\omega^2}\right)\right] \\ \label{eq:Delta_2f_NLO}
+  \left(\frac{\bar{w}_v}{\bar{w}_d}\right)^2 e^{-2\Gint(\tau_1 + \tau_2 - 2 \tini)} \cos \left[(\tau_1-\tau_2) \left(k + m\omega \left(1 + \frac{\delta\Gamma^2}{2\omega^2}\right) + n\tilde\omega\left(1+\frac{\delta\Gamma^2}{2\tilde\omega^2}\right)\right)\right]
 \biggr\}\, ,
\end{align}
where $\omega = c_s p$, $\tilde{\omega} = c_s \tilde{p}$. Let us note that this formula contains also the leading order result, i.e., in the limit $\frac{\delta\Gamma^2}{2\omega^2},\frac{\delta\Gamma^2}{2\tilde\omega^2} \to 0$, formula~\eqref{eq:Delta_2f} is obtained after using trigonometric identities. Second, the correction to the amplitude of the damped mode also leads to the modification of the kinetic spectrum, consequently, we present here the correction to $E_\Pi$ 
\begin{align}\nonumber
 E_\Pi^{\NLO}(\tau_1,\tau_2,k) =  \frac{2k^2 \bar{w}_d^2}{\bar{\rho}^2}  \int_{-1}^1 \dd z \int_0^\infty \dd p \frac{p^2}{\tilde{p}^4}(1-z^2)^2\, \Biggl\{ \cos \omega (\tau_1-\tau_2)\cos \tilde{\omega} (\tau_1-\tau_2) 
\biggl[ E_\kin (p) E_{\kin,1}^{\NLO} (\tilde{p})+  E_\kin (\tilde{p})E_{\kin,1}^{\NLO} (p) \biggr] F^\NLO_1(\tau_1,\tau_2) \\
 \label{eq:EpiNLO}
 + \biggl[\cos \omega (\tau_1-\tau_2)\sin \tilde\omega (\tau_1-\tau_2) E_\kin (p) E_{\kin,2}^\NLO(\tilde{p}) + \cos \tilde\omega (\tau_1-\tau_2)\sin \omega (\tau_1-\tau_2) E_\kin (\tilde{p}) E_{\kin,2}^\NLO (p)\biggr]F_2^\NLO(\tau_1,\tau_2) \Biggr\}\, ,
\end{align}
\end{strip}
\leavevmode\hspace*{-\parindent}%
where 
\begin{align}\nonumber
    F^\NLO_1(\tau_1,\tau_2) = \frac{\bar{w}_d  \bar{w}_v}{ (\bar{w}_d + \bar{w}_v)^2 } & \Bigl[e^{-2\Gint(\tau_1-\tini)} + e^{-2\Gint(\tau_2-\tini) }\\ 
    &+2\frac{\bar{w}_v}{\bar{w}_d}e^{-2\Gint(\tau_1+\tau_2-2\tini)}\Bigr]\,,\\
    F^\NLO_2(\tau_1,\tau_2) = \frac{\bar{w}_d  \bar{w}_v}{ (\bar{w}_d + \bar{w}_v)^2 } & \left[e^{-2\Gint(\tau_2-\tini)} - e^{-2\Gint(\tau_1-\tini) } \right]\,,
\end{align}
$E_\kin$ was defined in \cref{eq:Ekink}, and
\begin{equation}
    E_{\kin,n}^\NLO (k) =  \frac{\Gamma_V}{3 c_s k}\, \frac{k^2}{2\pi^2 R_\ast^3 \beta^6} \int \dd\tilde{T} \tilde{T}^6 \nu(\tilde{T}) \E_n^\NLO (k \tilde{T}/\beta)\,
\end{equation}
with
\begin{align}
    \E^\NLO_1 (\chi) = \frac{1}{2} c_s f'(\chi) l(\chi)\,,  \quad
    \E^\NLO_2 (\chi) = \frac{1}{2} f'(\chi)^2\,.
\end{align}
As expected, $E_\Pi^{\NLO}(\tau_1,\tau_2,k) $ vanishes in the limit $\frac{\Gamma_V}{c_s k} \to 0$. Note that in \cref{eq:Delta_2f_NLO,eq:EpiNLO}, the terms with $\sin$ or $\cos$ functions with arguments proportional to $\tau_1 + \tau_2 - 2\tini$ are neglected similarly as in the LO result~\eqref{eq:omGWfinal}, since these terms are expected to average out after integration over momenta.

We evaluated the GW spectra taking into account the corrections from \cref{eq:Delta_2f_NLO,eq:EpiNLO}, and assuming $\varkappa^{\M}_{\V} \sim \varkappa^{\M}_{\T}$, i.e., $|\delta\Gamma| \sim \Gint/2$ and $\Gamma_V \sim 3/4 \Gint$, see the \texttt{Mathematica} code~\cite{git_SSModel}.
For both types of correction, the plots for the parameter points discussed in the main text were not modified in a visible way compared to the LO result.
Let us note that for $\omega \leq |\delta\Gamma|$, the relative mode becomes overdamped, see \cref{eq:DampedModes}. This would change the calculation of the GW spectrum, however, we do not expect an effect on the results presented in the main text. A more detailed consideration of the $k\to 0 $ limit is left for future work.

\bibliographystyle{elsarticle-num} 
\bibliography{references}

\end{document}